\documentclass[epj, nopacs]{svjour}
\usepackage{amsmath,amssymb,graphicx,bm}
\usepackage{color}
\usepackage[bottom]{footmisc}

\newcommand{\beq}{\begin{equation}}
\newcommand{\eeq}{\end{equation}}
\newcommand{\bea}{\begin{eqnarray}}
\newcommand{\eea}{\end{eqnarray}}
\newcommand{\ba}{\begin{align}}
\newcommand{\ea}{\end{align}}
\newcommand{\bfig}{\begin{figure}}
\newcommand{\efig}{\end{figure}}

\newcommand{\D}{\displaystyle}

\newcommand{\tplus}{t_{+}}

\begin{document}

\title{Implications of unitarity and analyticity for the $D\pi$  form factors}
\author{B.\ Ananthanarayan$^{a}$  \and 
I. Caprini$^{b}$  \and I. Sentitemsu Imsong$^{a}$}
\institute{$^a$ Centre for High Energy Physics,
Indian Institute of 
Science, Bangalore 560 012, India \\
$^b$ Horia Hulubei National Institute for Physics and Nuclear Engineering,
P.O.B. MG-6, 077125 Magurele, Romania }


\abstract{We consider the vector and scalar form factors of the charm-changing current responsible 
for the semileptonic decay $D\rightarrow \pi l \nu$.   Using as input
dispersion relations and unitarity for the moments of suitable heavy-light correlators evaluated 
with Operator Product Expansions, including $O(\alpha_s^2)$ terms  in perturbative QCD, we constrain
the shape parameters of the form factors and find exclusion regions for zeros on the real axis and in the
complex plane.  For the scalar form factor, a low energy theorem and phase information on the unitarity 
cut are also implemented  to further constrain the shape parameters. 
We finally propose new analytic expressions for the $D\pi$ form factors, derive constraints on the  relevant coefficients from unitarity and analyticity, and briefly discuss the usefulness of the new parametrizations for describing semileptonic data.}

\titlerunning{D$\pi$ form factor}

\authorrunning{B.Ananthanarayan et al.} 

\maketitle


\section{Introduction} \label{sec:intro}
The charm-changing current responsible for the
semileptonic decay $D\rightarrow \pi l \nu$ is characterized by the vector and scalar 
form factors $f_+(t)$ and $f_0(t)$, defined by
\begin{multline}\label{eq:matrix}
\langle \pi^-(p') | \overline{d}\gamma_\mu c |D^0(p) \rangle  = (p'+p)_\mu f_+(t)+(p-p')_\mu f_-(t),
\end{multline}
\beq\label{eq:f0}
f_0(t)=f_+(t)+\frac{t}{M_D^2-M_\pi^2} f_-(t), \quad t=q^2=(p-p')^2.
\eeq
In the isospin limit the $D^+\to\pi^0$ form factors are obtained from $f_\pm(t)$ by multiplying with $1/\sqrt{2}$.

 The knowledge of the shape of the $D\pi$ form factors in the physical region 
$M_l^2\leq t\leq (M_D-M_\pi)^2$ is of interest for the determination of the element $|V_{cd}|$ of the Cabibbo-Kobayashi-Maskawa (CKM) matrix entering precision tests of the Standard Model. Recent measurements of the branching fractions of the semileptonic decays  $D\rightarrow \pi l \nu$ and $D\rightarrow K l \nu$ by the CLEO collaboration  \cite{Ge:2008yi,Besson:2009uv} renewed the interest in the theoretical study of these processes.

Earlier studies on  the heavy-light form factors are based on simple pole parametrizations  which implement  heavy quark scaling laws   \cite{Becirevic:1999kt}.
 The charm-changing form factors have been studied also in phenomenological
models based on heavy quark and chiral symmetry \cite{Fajfer:2004mv}, and more recently in the 
framework of Quantum Chromodynamics (QCD) light-cone sum rules (LCSR) \cite{Khodjamirian:2009ys}.
Lattice studies have  been carried out for the $D\rightarrow \pi, K$ 
form factors in the whole physical region \cite{arXiv:0903.1664,Bailey:2010vz,Na:2010uf,DiVita:2011py}. 
The $D\rightarrow \pi l\nu$ decay has also been considered recently
in the emerging field of hard 
pion effective theory  \cite{Bijnens:2010ws}.

Analyticity and unitarity are useful tools for improving the knowledge 
on the form factors. 
The standard dispersion relations are of little use in the present case due to the
scarce information on the $D\pi$ form factors on the unitarity 
cut.  However, the method of unitarity 
bounds \cite{Okubo,SiRa} proves to be a useful approach in cases
such as this. The method compensates for the lack of experimental information on 
the cut by an upper bound on the modulus squared of the form factor, 
obtained from unitarity and a dispersion relation for a suitable 
correlator of the same current, which can be evaluated by 
Operator Product Expansion (OPE) in the spacelike region.
Employing standard mathematical techniques, one can then 
correlate the values of the form factor and its derivatives at 
different points inside the analyticity domain. 
Various versions of the method were applied to the pion 
electromagnetic form factor 
\cite{Caprini2000,Ananthanarayan:2008qj,Ananthanarayan:2008mg,Abbas:2009ye,Anant:2011}, 
the $K\pi$ form factors \cite{MM,AES,BoMaRa,BC,Hill,Abbas:2009uw,Abbas:2009dz,Abbas:2010jc,Abbas:2010ns}, 
as well as to the heavy-heavy \cite{RaTa,deRafael:1993ib,Boyd:1995sq,Boyd:1995xq,Boyd:1997kz,CaLeNe} 
and heavy-light form factors \cite{Boyd:1994tt,Lellouch:1995yv,BoCaLe}.  
A review of the method was presented recently in \cite{Abbas:2010jc}.

 The $D\rightarrow \pi, K$ form factors  were investigated with this method 
some time ago \cite{Boyd:1994tt}. In this paper we revisit the problem, 
motivated in part by the progress in perturbative QCD calculations, 
which yield now the heavy-light correlators of interest to order $\alpha_s^2$  
\cite{Chetyrkin:2001je}. Since the correlators are given in \cite{Chetyrkin:2001je} 
only for a massless light quark, we restrict our study to the $D\rightarrow \pi$ form factors.  

In sect.  \ref{sec:method}  we briefly review the method of unitarity bounds in the heavy-light 
sector, extending previous studies by exploiting also higher moments of the corresponding 
correlation functions calculated in perturbative QCD \cite{Chetyrkin:2001je}.  The simultaneous 
use of several independent constraints is expected to increase the strength of the predictions. 
In sect.  \ref{sec:method}, we show also how to incorporate additional information on the form factors in 
the analyticity domain or on the unitarity cut. The numerical input of the method
is reviewed in sect. \ref{sec:input}.

In sect. \ref{sec:standard} we investigate the model-independent constraints imposed by analyticity and unitarity
on the low energy behaviour of the form factors. Specifically, we consider the shape parameters entering the Taylor expansion around $t=0$,
\beq
       f_{k}(t) = f_k(0)\left(1 +\lambda_{k}'\frac{t}{M_\pi^2} + \frac{1}{2} \lambda_{k}''\frac{t^2}{M_\pi^4} + \cdots\right), ~k=+,0,
       \label{eq:taylor}
\eeq 
and derive allowed ranges for the slopes  $\lambda_{k}'$ and the curvatures $\lambda_{k}''$.
We work with dimensionless parameters, the choice of $M_\pi^2$ as normalization scale  being merely a convention which does not influence the results (other choices, for instance $M_D^2$,  would simply scale the coefficients by a constant factor). In the same section we further apply the same formalism in order to isolate regions on the 
real $t$-axis and in the complex $t$-plane where zeros of the form factors  are excluded. The knowledge of the possible zeros is of interest, for instance, for the dispersive methods based on phase (Omn\`es-type representations) and for testing specific models of the form factors.  

In sect. \ref{sec:newparam} we propose a new parametrization of the $D\pi$ form factors, which generalizes the systematic expansion proposed in \cite{BoCaLe} for the $B\pi$  case by properly taking into account the position of the singularities generated by the first charm excited states. We derive constraints imposed by analyticity and unitarity on the coefficients of the parametrization and demonstrate its usefulness by fitting a sample of data points generated from the CLEO experimental results \cite{Ge:2008yi}.   Finally, in sect. \ref{sec:conc} we summarize our conclusions.
\section{Outline of the method}\label{sec:method}
 We start with the heavy-light invariant amplitudes $\Pi_+(q^2)$ and $\Pi_0(q^2)$ 
defined by the vector-vector correlation function
\begin{multline}\label{eq:corr}
-(q^2 g^{\mu\nu} - q^\mu q^\nu)\Pi_+(q^2) + q^\mu q^\nu \Pi_0(q^2) \\ 
= i\int d^4x e^{iqx} \langle 0|TV^\mu(x) V^{\nu\dag}(0)|0 \rangle 
\end{multline}
where $V_\mu=\bar d\gamma_\mu c$. It is convenient to consider
the moments of the invariant amplitudes at $q^2=0$, 
\begin{equation}\label{eq:chin}
\chi_{k}^{(n)} \equiv \frac{1}{n!}\frac{d^{n} }{(dq^2)^n}\left[\Pi_{k}(q^2) \right]_{q^2=0}, \quad k=+,0, \end{equation}
which satisfy dispersion relations of the form
\begin{equation}\label{eq:chindr}
\chi_{k}^{(n)} =\frac {1}{\pi} \int_{t_+}^\infty\! dt\,\frac{{\rm Im}\Pi_k (t+i\epsilon)}{t^{n+1} },  \quad k=+,0. 
\end{equation}
where $t_\pm=(M_D \pm M_\pi)^2$.

From QCD it is known that the amplitude $\Pi_+(q^2)$ satisfies a once subtracted dispersion 
relation, while for  $\Pi_0(q^2)$ an unsubtracted relation converges.  Therefore, 
the quantities  $\chi_{+}^{(n)}$  and $\chi_{0}^{(n)}$ are defined for $n\ge 1$ and  $n\ge 0$, respectively. 

The connection with the form factors 
$f_+(t)$ and $f_0(t)$  defined above is provided by unitarity: including in the unitarity 
sum for the spectral functions the contribution of the $D\pi$ states in the isospin limit 
leads to the inequalities 
\beq\label{eq:unit+}
{\rm Im}\Pi_+(t+i\epsilon) \geq \frac{3}{ 2}\frac{1}{ 48 \pi} \D \frac {\left[(t-t_+)(t-t_-)
\right]^{3/2}}{t^3} |f_+(t)|^2,
\eeq
\beq\label{eq:unit0}
{\rm Im} \Pi_0(t+i\epsilon) \ge \frac{3}{2} \frac{t_+ t_-}{ 16\pi}
\frac{[(t-t_+)(t-t_-)]^{1/2}}{ t^3} |f_0(t)|^2,
\eeq
 which hold for $t>t_+$. 

On the other hand,  $\chi_{k}^{(n)}$ can be calculated in OPE as the sum of 
perturbative ($PT$) and nonperturbative ($NP$) contributions:
\beq\label{eq:chiOPE}
\chi_{k}^{(n)} = \chi_k^{(n) PT} + \chi_k^{(n)NP}.
\eeq
 The perturbative parts of the moments of heavy-light correlators for $n\leq 7$ were 
calculated  up to two loops in \cite{Chetyrkin:2001je}.  Using these results we can write
\beq\label{eq:chiPT}
\chi_k^{(n)PT} = c_{k,0}^{(n)}+  c_{k,1}^{(n)}\, \alpha_s + c_{k,2}^{(n)} \, \alpha_s^2,
\eeq
where $\alpha_s$ is the strong coupling. The coefficients $c_{k,i}^{(n)}$ obtained using eqns. (34), (35) and the Appendix of \cite{Chetyrkin:2001je} are  compiled in Table \ref{table:pertcoeff} for several moments that will be used in this work.

The leading nonperturbative contribution of the quark and gluon condensates can be obtained from \cite{Lellouch:1995yv,Generalis}, and is written as
\beq\label{eq:chiNP+}
\chi_+^{(n)NP} = -\frac{1}{m_{c}^{2 (n+2)}} \left[\bar m_c\langle \bar u u \rangle
+ \frac{\langle\alpha G^{2} \rangle } {12\pi} \right],
\eeq
\beq\label{eq:chiNP0}
\chi_0^{(n)NP} = -\chi_+^{(n)NP}.
\eeq

\begin{table*}
\begin{center}
\caption{Perturbative coefficients $c_{k,j}^{(n)}$  defined in (\ref{eq:chiPT}) for the moments $\chi_k^{(n)}$ considered in the present work.}
\label{table:pertcoeff}
\begin{tabular}{llll}
\noalign{\smallskip}\hline
~~$i$	&  ~~~~~0 		&~~~~~1		&~~~~~2 \\
\noalign{\smallskip}\hline\noalign{\smallskip} 
$c_{+,i}^{(1)}$			&0.0024536			&0.0027977	&0.0066398	\\\vspace{0.05cm}
$c_{+,i}^{(2)}$			&0.0001690			&0.0002674 	&0.0009104	\\
$c_{+,i}^{(3)}$			&0.0000182			&0.0000342 	&0.0001371 	\\
\noalign{\smallskip}\hline\noalign{\smallskip} 
$c_{0,i}^{(0)}$			&0.0126651 			&0.0095139 	&0.0045948	\\\vspace{0.05cm}
$c_{0,i}^{(1)}$			&0.0008179 			&0.0013954 	&0.0037717	\\
$c_{0,i}^{(2)}$			&0.0000845			&0.0001860	&0.0006679 	\\
\noalign{\smallskip}\hline\noalign{\smallskip} 
\end{tabular}
\end{center}
\end{table*}
\begin{table*}
\begin{center}
\caption{Input values in the OPE calculation of the moments. }
\label{table:inputs}
\begin{tabular}{lll}
\noalign{\smallskip}\hline
\hspace{-0.2cm}Quantity 		&  Values	\\	
\noalign{\smallskip}\hline\noalign{\smallskip} 
$m_{c}^{pole}$ 			        &$1.968\, {\rm GeV} $	   \\    
$\overline{m}_{c}(\overline{m}_{c})$ 	&$(1.27_{-0.09}^{+0.07}) \,\rm GeV$	 \\ 
$\overline{m}_{c}(2\, {\rm GeV})$  		&$1.078\, {\rm GeV}$		\\
$\langle \overline{u}u \rangle (2 \,{\rm GeV})$  &$((-0.254 \pm 0.15)\, \rm GeV)^{3}$  \\
$\langle \alpha G^{2}\rangle$           &$(7.0\pm 1.3)\times 10^{-2}$\\
$\alpha_{s} (\overline{m}_c)$ 		& $0.39$		 \\
\noalign{\smallskip}\hline
\end{tabular}
\end{center}
\end{table*}
\begin{table*}
\begin{center}
\caption{OPE predictions for the vector correlators $\chi_+^{(n)}$.  }
\label{table:coeff1}
\begin{tabular}{llll}
\hline
$n$	& ~~~~~$\chi_+^{(n)PT}$ 	 	&~~~~~ $\chi_+^{(n)NP}$  &~~~~~ $\chi_+^{(n)}$ \\
\hline
1		& 0.0045547		&0.0002723		&0.0048270	\\
2		& 0.0004118		&0.0000704		&0.0004821	\\
3		& 0.0000524		&0.0000182		&0.0000706	\\
\hline
\end{tabular}
\end{center}
\end{table*}

\begin{table*}
\begin{center}
\caption{OPE predictions for the scalar correlators $\chi_0^{(n)}$.}
\label{table:coeff2}
\begin{tabular}{llll}
\hline
$n$	&~~~~~$\chi_0^{(n)PT}$ 	&~~~~~$\chi_0^{(n)NP}$ &~~~~~ $\chi_0^{(n)}$\\
\hline 
0		&0.0170744		&-0.0010543		&0.0160201\\
1		&0.0019357		&-0.0002723		&0.0016634\\
2		&0.0002586		&-0.0000704		&0.0001883\\
\hline
\end{tabular}
\end{center}
\end{table*}
From the dispersion relations (\ref{eq:chindr}) and the unitarity conditions (\ref{eq:unit+}) and (\ref{eq:unit0}), it follows that each form factor $f_{k}(t)$ satisfies a set of integral inequalities written as
\beq
 \frac{1}{\pi} \int^{\infty}_{\tplus } dt\ \rho_k^{(n)}(t) |f_{k}(t)|^{2} \leq \chi_k^{(n)},\quad \quad k=+,0,
        \label{eq:I}
\eeq
where the  weights $\rho_k^{(n)}(t)$ are the product of $1/t^{n+1}$ with the phase space factors entering the unitarity relations.
We use now the fact that the form factors $f_{+}(t)$ and $f_{0}(t)$ are analytic 
functions in the complex $t$-plane cut along the real axis from  $t_+$ to $\infty$, and apply
standard techniques to derive from eq.(\ref{eq:I})  
constraints on their values, in particular on the shape parameters and on the regions  
in complex energy plane where zeros are excluded. First, the problem is brought to a canonical form by mapping the $t$-plane into the interior of a unit disk.  This is achieved by
the general conformal transformation
\begin{equation}\label{eq:conformal}
\tilde z(t, t_0) = \frac{\sqrt{t_{+}-t_0}-\sqrt {t_{+} -t} } {\sqrt {t_{+}-t_0 }+\sqrt {t_{+}-t}},
\end{equation}
which maps the complex $t$-plane cut along the real axis for $t\ge t_+$  onto the unit disk $|z|<1$ in the complex plane $z\equiv\tilde z(t, t_0)$, such that $\tilde z(t_+, t_0)=1$ and  $\tilde z(\infty, t_0)=-1$. The real parameter $t_0<t_+$ is arbitrary and denotes the point mapped onto the origin, $\tilde z(t_0,t_0)=0$.  In the new variable, the inequality (\ref{eq:I}) takes the form 
\beq\label{eq:gI}
\frac{1}{2 \pi} \int_{0}^{2\pi} {\rm d}\theta |g_k^{(n)}(e^{i\theta})|^2 
	\leq \chi_k^{(n)},
\eeq
where the analytic functions $g_k^{(n)}(z)$ are defined as
\beq\label{eq:gz}
g_k^{(n)}(z) = f_k(\tilde t(z, t_0)) \,w_{k}^{(n)}(z).	
\eeq
Here  $\tilde t(z, t_0)=t_+-(t_+-t_0) (1-z)^2 /(1+z)^2$ is the inverse of the function defined in (\ref{eq:conformal}) and  $w_{k}^{(n)}(z)$ are
outer functions, {\it i.e.}  analytic and without zeros in
$|z|<1$, such that their modulus squared on the boundary, $z={\rm e}^{i\theta}$, is equal to $\rho_{k}^{(n)}(\tilde t({\rm e}^{i\theta}, t_0))$ multiplied by the Jacobian of the transformation (\ref{eq:conformal}). 
In our case the outer functions can be written in a compact form as
\begin{multline}\label{eq:vectorOF}
w_+^{(n)}(z)= \frac{1}{4 \sqrt{ 2\pi }}[t_+- \tilde t(z,t_0)] (t_+-t_0)^{-1/4} \\\times\left[\sqrt{t_+-t_-}+ \sqrt{t_+-\tilde t(z,t_0) }\right]^{3/2}\\\times\left[\sqrt{t_+-t_0}+ \sqrt{t_+-\tilde t(z,t_0) }\right]  \\\times  \left[\sqrt{t_+}+ \sqrt{t_+- \tilde t(z,t_0)}\right]^{-(n+4)}
\end{multline}
\beq\label{eq:scalarOF}
 w_{0}^{(n)}(z) =\frac{\sqrt{3}(m_{D}^2 - m_{\pi}^2) \, w_+^{(n)}(z)}{\sqrt{t_+-\tilde t(z,t_0)}  [\sqrt{t_+-t_-}+ \sqrt{t_+-\tilde t(z,t_0) }]}.
\eeq
In the notation we omitted for simplicity the dependence on $t_0$ of the functions $g_k^{(n)}(z)$,  $w_{+}^{(n)}(z)$ and  $w_{0}^{(n)}(z)$. 

 The analytic functions $g_{k}^{(n)}(z)$ admit the expansions
\beq\label{eq:gzseries}
g_{k}^{(n)}(z) = g_{k,0}^{(n)} + g_{k,1}^{(n)} z + g_{k,2}^{(n)} z^{2} + \cdots	
\eeq
convergent in $|z|<1$. From eq.(\ref{eq:gI}) it follows that the coefficients satisfy the inequality
\beq\label{eq:giI}
\sum\limits_{j=0}^{\infty}(g_{k,j}^{(n)})^2 \leq \chi_k^{(n)},\quad k=+,0.
\eeq
Since each term in the left side is positive, the largest domain allowed for the first Taylor coefficients $g_{k,j}^{(n)}$, $0\leq j\leq J-1$ is obtained from (\ref{eq:giI}) by setting the higher terms to zero.  More generally, a rigorous correlation between these  coefficients  and the values $\xi_{k,p}^{(n)}\equiv g_{k}^{(n)}(z_p)$ at some real points  $z_p$, $1\leq p\leq P$, is given by the determinantal inequality \cite{Okubo,SiRa,BoMaRa,Abbas:2010jc}
\beq\label{eq:determinant}
\left|
	\begin{array}{c c c c c c}
	\bar{\chi}_{k}^{(n)}  & \bar{\xi}_{k,1}^{(n)} & \bar{\xi}_{k,2}^{(n)}  & \cdots & \bar{\xi}_{k,P}^{(n)} \\	
	\bar{\xi}_{k,1}^{(n)} & \D \frac{z^{2J}_{1}}{1-z^{2}_1} & \D
\frac{(z_1z_2)^J}{1-z_1z_2} & \cdots & \D \frac{(z_1z_P)^K}{1-z_1z_P} \\
	\bar{\xi}_{k,2}^{(n)}  & \D \frac{(z_1 z_2)^{K}}{1-z_1 z_2} & 
\D \frac{(z_2)^{2J}}{1-z_2^2} &  \cdots & \D \frac{(z_2z_P)^K}{1-z_2z_P} \\
	\vdots & \vdots & \vdots & \vdots &  \vdots \\
	\bar{\xi}_{k,P}^{(n)} & \D \frac{(z_1 z_P)^J}{1-z_1 z_P} & 
\D \frac{(z_2 z_P)^J}{1-z_2 z_P} & \cdots & \D \frac{z_P^{2J}}{1-z_P^2} \\
	\end{array}\right| \ge 0.
\eeq
where
\beq\label{eq:tildeI}
\bar{\chi}_{k}^{(n)} = \chi_{k}^{(n)} - \sum_{j = 0}^{J-1} (g_{k, j}^{(n)})^2,
\eeq
and
\beq
\bar{\xi}_{k,p}^{(n)} = \xi_{k,p}^{(n)}- \sum_{j=0}^{J-1} g_{k, j}^{(n)} z_p^j, \quad p=1,2,...P.
\eeq
The  condition (\ref{eq:determinant}) is expressed in a straightforward way in terms of the values  $f_k(t_p)$ of
the form factors at  $t_p=\tilde t(z_p,t_0)$ and the derivatives at $t=0$, using eqns.
(\ref{eq:conformal}) and (\ref{eq:gz}).  The generalization to complex points $z_p$ can be found in \cite{BoMaRa,Abbas:2010jc}. 

In the present work we use the inequality (\ref{eq:determinant}) to obtain bounds on the slopes 
and curvatures of the form factors defined in eq.(\ref{eq:taylor}). As input  we shall take 
the values $f_+(0)= f_0(0)$ at $t=0$, which are known from LCSR \cite{Khodjamirian:2009ys}.  An additional piece of information for the scalar form factor is provided by a 
low-energy soft-pion theorem of the Callan-Treiman \cite{CallanTreiman} type, 
proved in \cite{DoKoSc}, which in the $D\pi$ case reads
\beq\label{eq:CT}
f_0(\Delta_{D\pi}) = f_D/f_\pi,
\eeq
where  $\Delta_{D\pi}= M_D^2-M_{\pi}^2$ is the relevant Callan-Treiman point  
and $f_D$ and $f_\pi$ are the meson decay constants.

 As shown in \cite{Abbas:2010jc}, 
from  eq.(\ref{eq:determinant}) one can derive also regions where zeros of the form factors are excluded: this 
is done in a straightforward way by including in eq.(\ref{eq:determinant})  the input $f_k(z_0)=0$ and finding the values $z_0$ for which the inequality is violated. 

The constraints on the shape parameters can be further improved if some 
information on the form factors on the unitarity cut is available, 
in particular, if the phase  $\delta_k(t)$  defined as 
\beq\label{eq:phase}
f_k(t+i\epsilon)=|f_k(t)| {\rm e}^{i \delta_k(t)}, ~~k=+,0,
\eeq
is known along a low-energy interval, $t_+\leq t\leq t_{in}$. The  implementation of this information 
can be done by the technique of generalized Lagrange multipliers and 
involves the  solution of an integral equation, applied first to the $K\pi$ form factors in \cite{MM,AES}.  A review of the method and more references are given in  \cite{Abbas:2010jc}. 

Previous work on unitarity constraints for the heavy-light form factors 
\cite{Boyd:1994tt,Lellouch:1995yv,BoCaLe},  exploited  
dispersion relations for the lowest  moments of the correlators, 
corresponding to $n=1$ ($n=0$)  for the vector (scalar) form factor. Here we use also the higher moments calculated to two loops in perturbative QCD \cite{Chetyrkin:2001je}. Specifically, we use the dispersion relations 
for the moments $\chi_+^{(1)},  \chi_+^{(2)}, \chi_+^{(3)}$ and $\chi_0^{(0)},  \chi_0^{(1)}, \chi_0^{(2)}$. From the inequalities (\ref{eq:gI}) we 
obtain, for each form factor,  a family of three different constraints. 
The final allowed domain for the parameters of interest will be 
the intersection of the three individual domains.

Up to now we did not specify the parameter $t_0$ appearing in the conformal mapping (\ref{eq:conformal}). 
One can show that the bounds presented above are independent on $t_0$: indeed, 
the bounds are obtained by solving extremum problems upon the class of analytic admissible 
functions satisfying an $L^2$ norm condition like (\ref{eq:gI}). Changing $t_0$ amounts to mapping 
a unit disk to another, and by this the class of admissible functions is not changed \cite{Duren}. 
For deriving the bounds reported in sect. \ref{sec:standard} we worked with $t_0=0$.

\section{Choice of the input \label{sec:input}}

We have presented in Table  \ref{table:pertcoeff}  
the coefficients $c_{k,j}^{(n)}$ from \cite{Chetyrkin:2001je}, 
entering the perturbative calculation of the moments $\chi_k^{(n)}$. 
In Table \ref{table:inputs} we compile other quantities entering
as input in the calculation of the moments. We exploited the approximate scale invariance of the product $ m_c \langle \overline{u}u \rangle$  and  evaluated it at a scale of 2 GeV. 
We took the masses and the strong
coupling constant from the PDG tables \cite{PDG} 
and used the renormalization group equations to evolve 
them to the relevant scale. 
The gluon condensate $\langle \alpha G^{2}\rangle$ has been taken 
from \cite{Narison:2010wb,Narison:2011xe}. The denominator 
of eq.(\ref{eq:chiNP+})  involves the pole mass \cite{Generalis}.
The $PT$ and $NP$ contributions and the total OPE predictions for $\chi_{+}^{(n)}$ and $\chi_{0}^{(n)}$ are summarised in Tables \ref{table:coeff1} and \ref{table:coeff2} for the values of $n$  specified above.

As mentioned earlier, we work in the isospin limit,  taking for convenience for the $D$ meson the
 mass  $M_D= 1.869 \,{\rm  GeV}$  of the neutral $D$ meson, and for $\pi$ the mass $M_\pi= 0.1395\,{\rm  GeV}$ of the charged pion. 
Then $t_+= 4.02\, {\rm  GeV}^2$ and $t_-= 2.98\, {\rm  GeV}^2$.  

In our analysis,  we used as input the value $f_+(0) = 0.67_{-0.07}^{+0.10}$ 
provided by LCSR \cite{Khodjamirian:2009ys}\footnote{Lattice simulations are in agreement with
this number: a relativisitic computation on a fine lattice
in the quenched approximation \cite{arXiv:0903.1664} led to the value $f_+(0)=0.74(6)(4)$,  
while the result  $f_+(0)= 0.65(6)(6)$ was obtained in  \cite{DiVita:2011py} using maximally twisted Wilson fermions with $N_f=2$.}, 
and $f_0(0)=f_+(0)$ as follows from eq.(\ref{eq:f0}). 
Using $f_D= (206.7 \pm 8.5 \pm 2.5)\, {\rm MeV} $ and $f_\pi=(130.41\pm0.03\pm 0.20)\,  {\rm MeV}  $ \cite{PDG},
the low-energy theorem (\ref{eq:CT}) gives  $f_0(\Delta_{D\pi})= 1.58 \pm 0.07$ as quoted in \cite{Khodjamirian:2009ys}.

In order to obtain the phase $\delta_k(t)$ defined by (\ref{eq:phase}), a method useful in the case of the pion electromagnetic or the $K\pi$ weak form factors is to 
invoke Fermi-Watson theorem  and take the precisely known phase shifts of the corresponding elastic partial waves. Since in the $D\pi$ case  
the elastic scattering is not yet investigated (except some comments on
 the $S$-wave in \cite{Bugg:2009tu}),  we shall roughly estimate
the phase of the form factors  from the masses and  widths of  the resonances dominant at low energies. 
 Namely, from the relativistic Breit-Wigner parametrization we obtain
\beq	\label{eq:phasedelta}
\delta(t) =  \arctan \left(\frac{M_R\Gamma(t)}{M_R^2-t}\right),
\eeq
where $M_R$ is the mass  and $\Gamma(t)$ the energy-dependent width  
\beq	\label{eq:gamma}
\Gamma(t) = \left(\frac{q(t)}{q(M_R^2)}\right)^{2 J+1} \frac{M_R}{\sqrt{t}}\,\Gamma_R,
\eeq
written in terms of the angular momentum  $J$, the width $\Gamma_R$ and the c.m. momentum $
q(t)= \sqrt{(t-t_-)(t-t_+)/4t}$.

The lowest vector $D^*$ and scalar $D_0^*$ excited states that couple to the $D\pi$ 
system produce singularities above the threshold, on the second Riemann sheet.  
The central values of the masses and widths of the lowest charged $D\pi$  vector and scalar resonances 
listed in  \cite{PDG} are: $M_{D^*} = 2010.25 \pm 0.14 \,{\rm  MeV}$, $\Gamma_{D^*} = 96 \pm 4 \pm 22 \,{\rm  MeV}$ 
and $M_{D_0^*}= 2403 \pm 14 \pm 35 \,{\rm  MeV}$,  $\Gamma_{D_0^*}= 283 \pm 24 \pm 34\, {\rm MeV}$.

\bfig[ht]	
	\begin{center}\vspace{0.cm}
	 \includegraphics[angle = 0, clip = true, width = 2.3 in]
{phase.eps}	
	\end{center}\vspace{-0.2cm}
	\caption{Phase of the $D\pi$ scalar form factor using a relativistic Breit-Wigner parametrization of the lowest resonance $D_{0}^*$.}  
	\label{fig:phase}
\efig
We note that the vector resonance $D^*$ is very close to the threshold, so that a reasonable expression of the phase cannot be obtained from a
Breit-Wigner parametrization, but in the scalar case we can assume that the 
phase $\delta_0(t)$ is reliably described by the expression (\ref{eq:phasedelta}) with $J=0$,  
which is plotted in Fig. \ref{fig:phase}.  In our work we shall implement the phase up 
to the point $t_{in}= (2.6 \,{\rm GeV})^2$,  close to the first inelastic $D\eta$ channel opening at 2.42 GeV.

\section{Constraints on the shape parameters and zeros\label{sec:standard}}

The interior of the ellipses shown in
Fig.\ref{fig:vec_slope_curv} represent the allowed domains in the slope-curvature plane for the 
vector form factor, obtained with  three moments $\chi^{(n)}_+$ of the vector correlator. 
We use the inequality (\ref{eq:determinant}) with no other input  on the cut or in the 
analyticity domain except the value of $f_+(0)$ given above. The best (smallest) domain 
is given by the lowest moment, but the higher moments contribute to slightly reducing it, since one must take the intersection of all the domains in order to fulfill simultaneously the constraints.
We also indicate the slope and curvature of the simple pole ansatz \cite{Becirevic:1999kt}
\beq	\label{eq:polevector}
f_{+}(t) = \frac{f_+(0)}{(1-t/M_{D^*}^2)(1-\alpha_{D\pi}t/M_{D^*}^2)},
\eeq
 with the parameters 
 $\alpha_{D\pi} = 0.21_{-0.07}^{+0.11}$ and $M_{D^*}= 2.007\,{\rm GeV}$ proposed in  \cite{Khodjamirian:2009ys}.
As seen from the figure, the point satisfies the unitarity constraints.

In the scalar case we use additional information on the phase and the soft pion theorem 
given by eq.(\ref{eq:CT}). Fig. \ref{fig:scalar_slope_curv}  illustrates the effect of  
these additional constraints, in the particular case of the bounds derived from 
the lowest moment $\chi_0^{(0)}$.
We also show the slope and curvature from the pole ansatz \cite{Becirevic:1999kt}
\beq	\label{eq:polescalar}
f_0(t) = \frac{f_+(0) }{1-t/(\beta_{D\pi}M_{D_0^*}^2)},
\eeq
with the parameters $\beta_{D\pi} = 1.41\pm 0.06\pm0.07$ and  $M_{D_0^*}= 2.318 \,{\rm GeV}$ suggested in \cite{Khodjamirian:2009ys}.
The point satisfies the constraint imposed by the lowest moment, taking into account also the phase and the low-energy theorem (\ref{eq:CT}).

\bfig[ht]	
	\begin{center}\vspace{0.cm}
	 \includegraphics[angle = 0, clip = true, width = 2.3 in]
{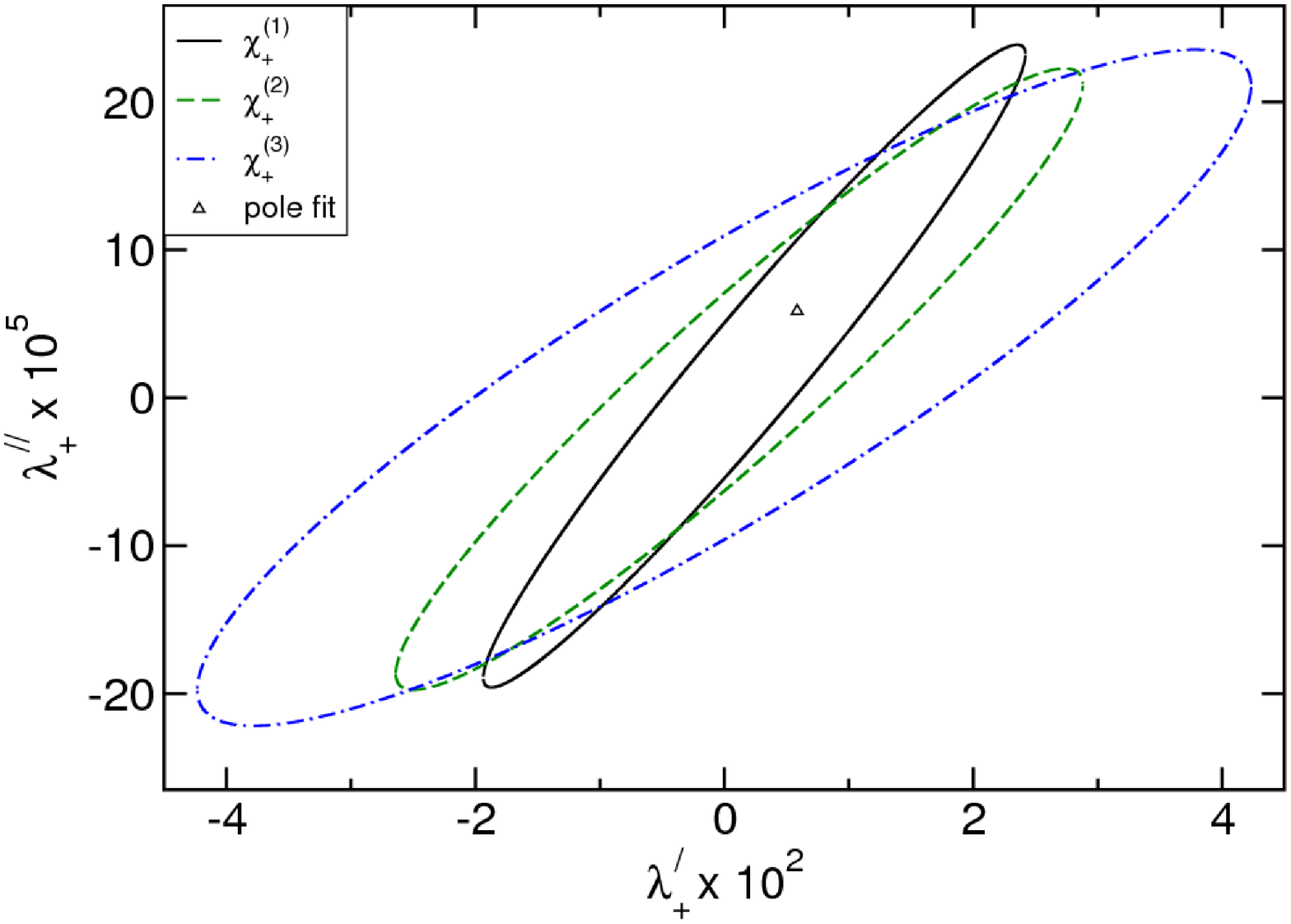}	
	\end{center}\vspace{-0.2cm}
	\caption{Constraints on the slope and curvature of the vector form factor obtained using as input different moments of the correlator. The point indicates the slope and curvature of the pole ansatz  (\ref{eq:polevector}).}
	\label{fig:vec_slope_curv}
\efig	

\bfig[ht]	
	\begin{center}\vspace{0.cm}
	 \includegraphics[angle = 0, clip = true, width = 2.3 in]
{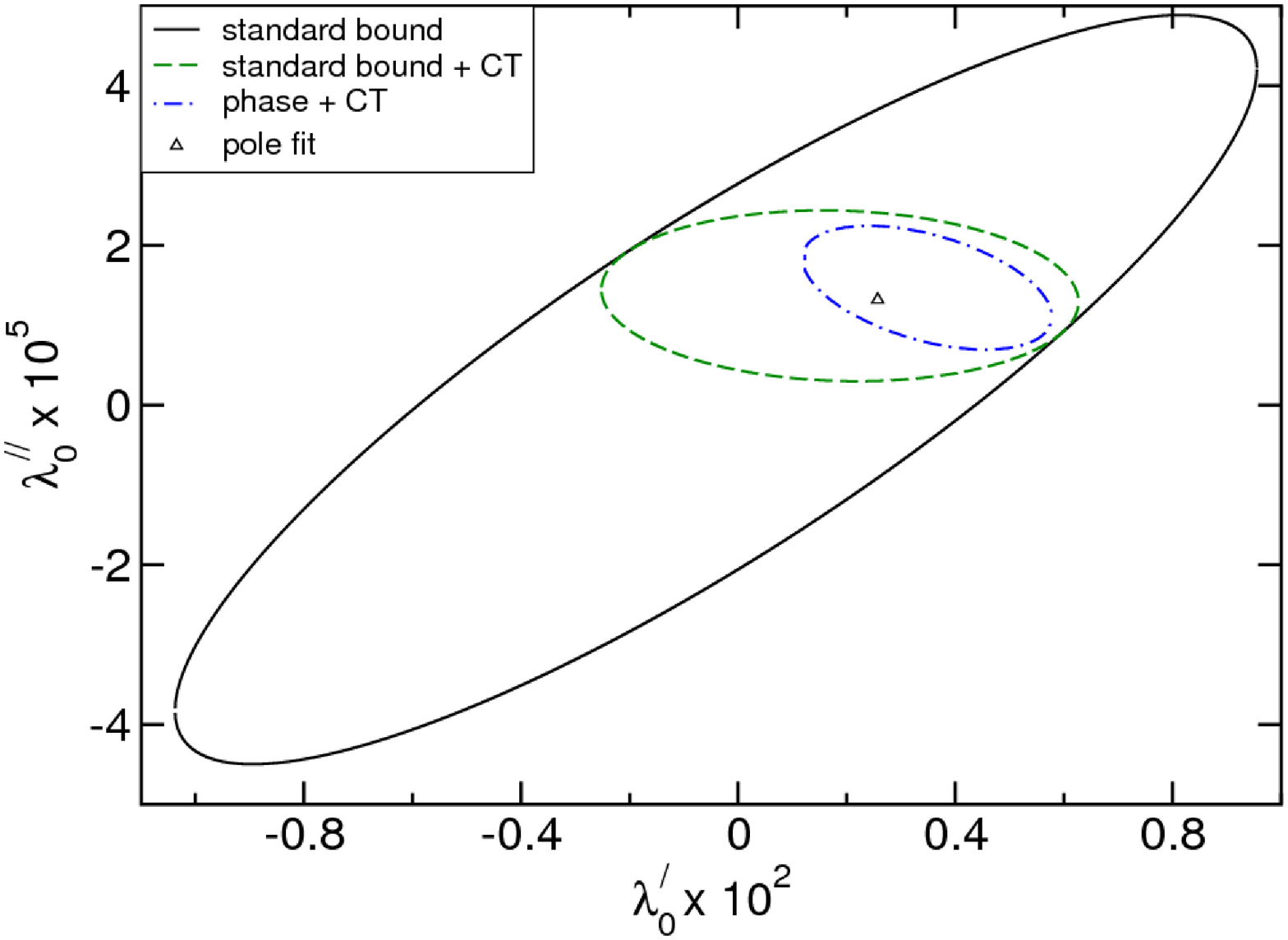}	
	\end{center}\vspace{-0.2cm}
	\caption{Constraints on the slope and curvature of the scalar form factor obtained with the moment $\chi_0^{(0)}$, from the standard bound and  by including the phase and the low-energy theorem (\ref{eq:CT}). 
The point indicates the slope and curvature of the pole ansatz (\ref{eq:polescalar}).}  
	\label{fig:scalar_slope_curv}
\efig	

In Fig. \ref{fig:scalar_moments} we show  the allowed 
domains obtained by using different moments $\chi_0^{(n)}$ of the scalar correlator. 
The  phase and the low-energy constraint (\ref{eq:CT}) have been included in all cases. Again, 
the smallest ellipse is obtained with the lowest moment  $\chi_0^{(0)}$, but 
the simultaneous constraints reduce further this domain.  We obtain rather  small allowed regions in the 
slope-curvature plane   when the phase as well as the low energy theorem are 
simultaneously taken into account. In particular, the point corresponding to the pole 
ansatz (\ref{eq:polescalar}) 
falls outside the allowed regions imposed by the moments with $n=1$ and $n=2$.  

\bfig[ht]	
	\begin{center}\vspace{0.cm}
	 \includegraphics[angle = 0, clip = true, width = 2.3 in]
{scalar_moments.eps}	
	\end{center}\vspace{-0.2cm}
	\caption{Constraints obtained by considering different moments of the scalar correlator, including information about the phase and the low-energy theorem (\ref{eq:CT}). The point indicates the slope and curvature of the pole ansatz (\ref{eq:polescalar}).}  
	\label{fig:scalar_moments}
\efig

\bfig[ht]	
	\begin{center}\vspace{0.3cm}
	 \includegraphics[angle = 0, clip = true, width = 2.3 in]
{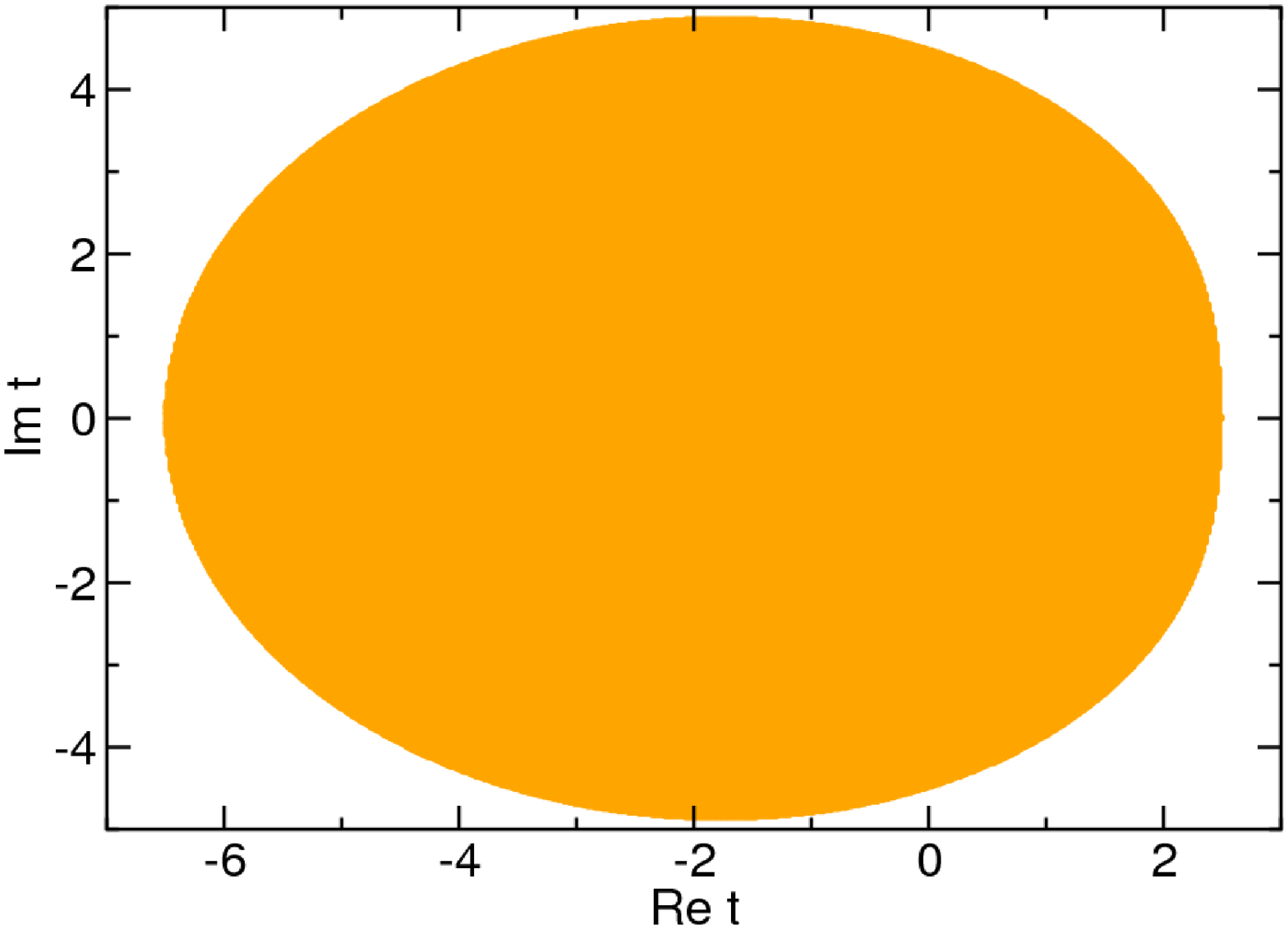}	
	\end{center}\vspace{-0.2cm}
	\caption{Domain without zeros for the vector form factor, obtained from the lowest moment $\chi_+^{(1)}$ and the input $f_+(0)$.}  
	\label{fig:vec_zeros}
\efig

\bfig[ht]	
	\begin{center}\vspace{0.3cm}
	 \includegraphics[angle = 0, clip = true, width = 2.3 in]
{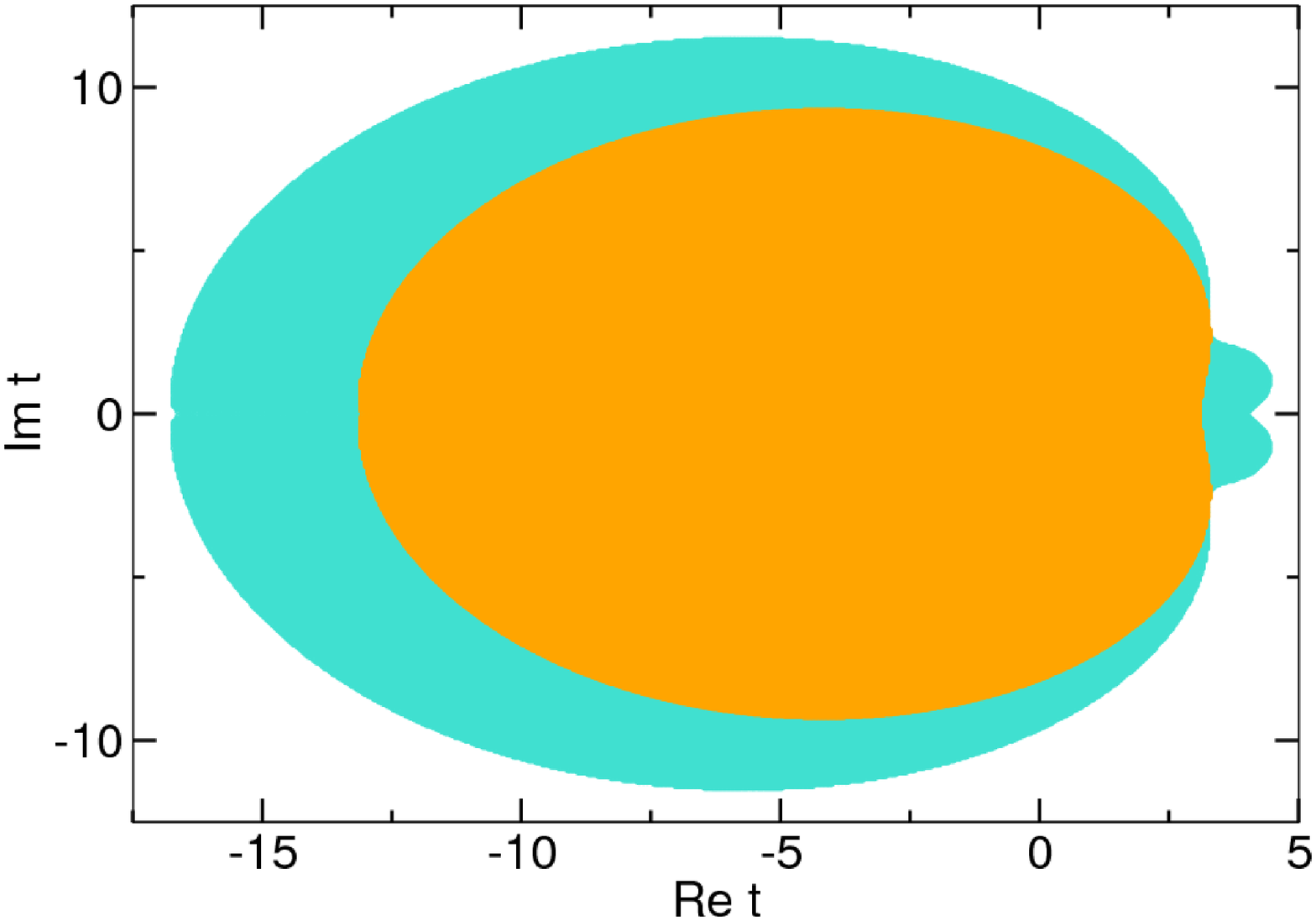}	
	\end{center}\vspace{-0.2cm}
	\caption{Domain without zeros for the scalar form factor, obtained from the lowest moment $\chi_0^{(0)}$ and the input $f_0(0)$ (smaller region) and using in addition the constraint (\ref{eq:CT}) (larger  region).}  
	\label{fig:scalar_zeros}
\efig	

We mention that the above bounds were obtained  with the central values of the input parameters. 
By varying simultaneouly all the input values we can obtain more conservative regions, which are slightly 
larger than the domains shown in the figures.  The modification of the moments $\chi_k^{(n)}$ entering the 
inequalities (\ref{eq:gI}) affects the results in a monotonous way: larger values of 
 $\chi_k^{(n)}$ lead to weaker bounds. In our study we reduced this source of uncertainty by 
using more precise perturbative values of heavy-light current correlators, calculated in  
\cite{Chetyrkin:2001je} at order $\alpha_s^2$.  We have also studied the influence of the uncertainties
in the resonance parameters for the mass and  width of the scalar resonance \cite{PDG}, 
and the  uncertainty on the form factor values at $t=0$ and at the Callan-Treiman point (\ref{eq:CT}).  
For instance, for the lowest moment illustrated in Fig.
\ref{fig:scalar_slope_curv}, the union of the ellipses resulting from the
variation of $f_+(0)$, $f_0(\Delta_{D\pi})$, $M_{D^*}$ and   
$\Gamma_{D^*}$  within the errors quoted in sect. \ref{sec:input}, and by
enlarging  $\chi_+^{(1)}$ by 10\% yield the allowed ranges 
$0.08\cdot 10^{-2} \lesssim \lambda_{0}' \lesssim 0.67\cdot 10^{-2}$
and $0.51\cdot 10^{-5} \lesssim \lambda_{0}'' \lesssim 2.56 \cdot 10^{-5}$, somewhat 
larger than ranges corresponding to the central values of the input,
 $0.12\cdot 10^{-2} \lesssim \lambda_{0}' \lesssim 0.58\cdot 10^{-2} $ 
and $0.69\cdot 10^{-5} \lesssim \lambda_{0}'' \lesssim 2.25\cdot 10^{-5} $.

A similar enlargement of the ellipses given by the next moments in Fig. \ref{fig:scalar_moments} makes the pole ansatz compatible with the allowed domain given by $\chi_1^{(0)}$.  It still remains  
outside the constraint yielded by $\chi_2^{(0)}$, but very close to the
edge of the allowed domain.

As discussed in earlier works \cite{Abbas:2010ns,Anant:2011},  
the knowledge of the zeros is useful for specific fits and for testing general ideas about form factors. Although
chiral symmetry for instance implies the existence of zeros in
scattering amplitudes, there is not much information theoretically
about them in the case of form factors.  Nevertheless, phenomenological
dispersive analyses often assume that zeros are absent in their fits
to experimental data.  Whereas remote zeros do not make any appreciable
effect on these, nearby zeros if present will affect the behaviour
of such fits and influence the number of subtractions. Therefore it
is important to exclude zeros in the low-energy region.

For illustration, we consider in the following only the constraints imposed by the lowest moments, corresponding to $n=1$ in the vector case and 
$n=0$ in the scalar one, using as input the quoted value of $f_+(0)$ from LCSR. In the vector case  we find that zeros on the 
real axis are excluded in the range $(-1.0,\, 0.80)\,{\rm GeV}^2$. For the scalar case   the range with  no zeros 
is $(-2.51,\, 1.55)\,{\rm GeV}^2$ for the standard bounds,  while with the inclusion of the low-energy constraint
  (\ref{eq:CT}) 
the range is increased to  $(-3.55,\, 3.89)\,{\rm GeV}^2$. In Figs. \ref{fig:vec_zeros} and \ref{fig:scalar_zeros}   
we present the regions of excluded zeros in the complex $t$-plane for the vector and
scalar form factors. In the scalar case the zeros are excluded in a larger domain if the 
low-energy constraint (\ref{eq:CT}) is also imposed.  In principle it would be possible
to include also the phase and solve the integral equations for each point
that is being tested.  While an improvement of the size of the
domains is expected, the computation is a laborious one and
is beyond the scope of the present investigation. 

\section{New parametrization of the $D\pi$ form factors}\label{sec:newparam}

Until recently,  the most popular parametrization of the heavy-light form factors  was the pole ansatz \cite{Becirevic:1999kt}, given in eqs. (\ref{eq:polevector}) and  (\ref{eq:polescalar}) .
A more general parametrization, based on a systematic expansion in powers of a conformal variable, was proposed in \cite{BoCaLe} for the $B\pi$ vector form factor. 
In  \cite{Khodjamirian:2009ys} the same type of parametrization was adopted also in the $D\pi$ case. 
 Specifically, the vector $D\pi$  form factor was written in \cite{Khodjamirian:2009ys} as 
\beq\label{eq:zK}
f_+(t)=\frac{1}{1-t/M_{D^*}^2}\,\sum\limits_{j=0}^J b_j z^j,
\eeq
where  $z=\tilde z(t, t_0)$ is the variable defined in (\ref{eq:conformal}).
However, this straightforward generalization is not entirely consistent, since the  pole 
due to the lowest $D^*$ resonance is above the threshold on the second Riemann sheet, while in the expression (\ref{eq:zK}) the singularity is on the real axis. Due to this fact, as we shall see, unitarity
 constraints on the free parameters $b_j$ cannot be 
derived. We mention that these constraints are useful not only for restricting the range of the independent parameters, but also for estimating the truncation error \cite{BoCaLe}.  

In the present work we write down improved parametrizations that take
into account the proper position of the singularities produced by the first charm excited states. We propose for the vector form factor, instead of eq. (\ref{eq:zK}), 
the representation\footnote{It is not necessary to use a  $P$-wave phase-space in the Breit-Wigner 
(BW) denominator, since the proper behaviour at threshold will be imposed below for the entire function $f_+(t)$.}
\beq\label{eq:zparam+}
f_+(t)=\frac{M_{D^*}^2}{M_{D^*}^2-t +\sqrt{1-t/t_+} M_{D^*} \Gamma_{D^*}}\,\sum\limits_{j=0}^J b_{+,j} z^j, 
\eeq
 where as above  $z=\tilde z(t, t_0)$ is the conformal mapping  (\ref{eq:conformal}). Similarly, for the scalar form factor, we write
\beq\label{eq:zparam0}
f_0(t)=\frac{M_{D_0^*}^2}{M_{D_0^*}^2-t +\sqrt{1-t/t_+} M_{D_0^*} \Gamma_{D_0^*}}\,\sum\limits_{j=0}^J b_{0,j} z^j.
\eeq
The coefficients $b_{k,j}$ are the parameters to be used in fits of data. Unitarity and analyticity 
imply that they are not completely free, but satisfy a constraint \cite{BoCaLe}. 
In order to derive it, we insert the representations (\ref{eq:zparam+})  and (\ref{eq:zparam0}) into the definition (\ref{eq:gz}) of the functions $g_k^{(n)}(z)$, extract the corresponding Taylor coefficients and apply the inequality (\ref{eq:giI}). For simplicity, we consider only the constraints obtained with the lowest moments of the correlators. Then the above steps lead to the inequalities
\beq\label{eq:unitbn}
\sum\limits_{i,j=0}^J B_{ij}^{(k)} b_{k, i} b_{k,j} \leq 1,\quad \quad k=+,0,
\eeq
where $B_{ij}^{(k)}$ are calculated as  \cite{BoCaLe}. 
\beq\label{eq:Bjk}
 B_{ij}^{(k)} =\frac{1}{\chi_k}  \sum\limits_{m=0}^\infty \eta_{k,m} \eta_{k,m+|i-j|}, \quad \quad k=+,0.
\eeq
Here  we used the notations $\chi_+\equiv \chi_+^{(1)}$, $\chi_0  \equiv\chi_0^{(0)}$, and the numbers $\eta_{k,m}$ are the Taylor coefficients appearing in the expansions
\beq\label{eq:eta+}
\frac{M_{D^*}^2\, w_+^{(1)}(z) }{M_{D^*}^2-\tilde t(z,t_0) + \sqrt{1- \tilde t(z,t_0)/t_+} M_{D^*} \Gamma_{D^*}} =  \sum\limits_{m=0}^\infty \eta_{+,m} z^m,
\eeq
\beq\label{eq:eta0}
 \frac{M_{D_0^*}^2 \,w_0^{(0)}(z)}{M_{D_0^*}^2-\tilde t(z,t_0) + \sqrt{1- \tilde t(z,t_0)/t_+}  M_{D_0^*} \Gamma_{D_0^*}} =  \sum\limits_{m=0}^\infty \eta_{0,m} z^m,
\eeq
where the outer functions $w_k^{(n)}(z)$ are  defined in eqs. (\ref{eq:vectorOF}) and (\ref{eq:scalarOF}), 
and  $\tilde t(z,t_0)$ is the inverse of (\ref{eq:conformal}). 

  As remarked in sect. \ref{sec:method}, the bounds derived in the previous section are 
independent of the choice of the parameter $t_0$ (in the calculations we have set  $t_0$ to 0). 
By contrast, the parametrizations given above are based on truncated expansions and depend on $t_0$. For $t_0=0$ the semileptonic region $(0, t_-)$ is mapped onto the range $(0, 0.35)$ in the $z$-plane. This may lead to quite large truncation errors at the right end of the physical region. As suggested first in \cite{Boyd:1995xq,Boyd:1994tt}, it is more convenient to choose  $t_0$ such as to map the physical range symmetrically around the origin of the $z$-plane, {\it i.e.} $\tilde z(0,t_0)=-\tilde z(t_-,t_0)$. This gives the optimal value
\beq\label{eq:t0opt}t_{opt}\equiv
(m_D+m_\pi)(\sqrt{m_D}-\sqrt{m_\pi})^2=1.97 \,\mbox{GeV}^2, \eeq 
when the physical range is mapped onto the segment (-0.17, 0.17) in the $z$-plane.  In the following we shall investigate both choices $t_0=0$ and $t_0=t_{opt}$.

Since the outer functions and the BW representations are analytic in the unit disk  and have no singularities on the boundary $|z|=1$
(the denominators vanish outside the unit disk),  the expansions in the right hand sides of eqns.(\ref{eq:eta+}) and (\ref{eq:eta0}) are rapidly convergent. Therefore the quantities  $B_{ij}^{(k)}$ can be calculated with precision by truncating the series in (\ref{eq:Bjk}) at a finite order.\footnote{A pole situated on the real axis, as in (\ref{eq:zK}), is mapped by (\ref{eq:conformal}) onto a point $z_0$ on the boundary of the unit disk, $|z_0|=1$.  The quantities $\eta_{k,m}$ involve the powers $1/z_0^m$, and it is easy to see that the series in the right side of (\ref{eq:Bjk}) is not convergent in that case.}

 The numbers  $B_{ij}^{(+)}$ and $B_{ij}^{(0)}$  up to  $J=5$  for $t_0=0$ are
\bea\label{eq:Bij+}
&& B_{00}^{(+)}=4.45\cdot 10^{-2},\quad B_{01}^{(+)}=2.53 \cdot 10^{-2}, \nonumber \\ 
&& B_{02}^{(+)}=-4.53\cdot 10^{-4},\quad
B_{03}^{(+)}=-3.60\cdot 10^{-3}, \nonumber \\ &&B_{04}^{(+)}=-4.60 \cdot 10^{-5},\quad B_{05}^{(+)}=-5.81\cdot 10^{-4},
\eea 
\bea\label{eq:Bij0}
&&B_{00}^{(0)}= 2.83,\quad B_{01}^{(0)}= 1.18 ,\quad B_{02}^{(0)}=-1.68  \nonumber \\
&&B_{03}^{(0)}=-2.29,\quad B_{04}^{(0)}=- 0.32,\quad B_{05}^{(0)}=1.60.
\eea 
while for $t_0=t_{opt}$ the values read
\bea\label{eq:Bijtopt+}
&& B_{00}^{(+)}=4.45\cdot 10^{-2},\quad B_{01}^{(+)}=1.69 \cdot 10^{-2}, \nonumber \\ 
&& B_{02}^{(+)}=-8.51\cdot 10^{-3},\quad
B_{03}^{(+)}=-8.87\cdot 10^{-4}, \nonumber \\ &&B_{04}^{(+)}=5.75 \cdot 10^{-4},\quad B_{05}^{(+)}=-3.10\cdot 10^{-3},
\eea 
\bea\label{eq:Bijtopt0}
&&B_{00}^{(0)}= 1.91,\quad B_{01}^{(0)}= 0.25 ,\quad B_{02}^{(0)}=-1.61  \nonumber \\
&&B_{03}^{(0)}=-0.53,\quad B_{04}^{(0)}=1.09,\quad B_{05}^{(0)}=0.51.
\eea 
\begin{table*}
\begin{center}
\caption{Sample of data on $f_+(t)$ used for a fit with the parametrization (\ref{eq:zparam+}).}
\label{table:coeff3}
\begin{tabular}{lllllllllll}
\hline
$n$&1&2&3&4&5&6&7&8&9&10\\\hline
$t_n$ & 0& 0.15& 0.45& 0.745& 1.04 & 1.34 & 1.59 & 1.89&  2.23 & 2.68		\\
$f_+(t_n)$ &0.67& 0.68 &0.74 & 0.760& 0.94& 0.98 &1.06 &
1.57& 1.69 & 1.94		\\
$\delta f_+(t_n)$&0.10& 0.03& 0.04& 0.05& 0.05 &0.06 &
0.08 & 0.11 & 0.23 & 0.29	\\
\hline
\end{tabular}
\end{center}
\end{table*}

The remaining   coefficients are obtained from the relations
  $B_{i(i+j)}^{(k)}=B_{0j}^{(k)}$ and the symmetry property
  $B_{ij}^{(k)}=B_{ji}^{(k)}$ which follow from (\ref{eq:Bjk}). In the calculation we used for the masses and widths of the resonances the central values from \cite{PDG}, quoted above. 
 The smaller values of $B_{ij}^{(+)}$ are due partly to the value of the  QCD vector correlator and the form of the unitarity inequality (\ref{eq:unit+}), and partly to the parameters of the $D^*$-resonance.

 Additional information, like the values of the form factors at $t=0$ and the low energy theorem (\ref{eq:CT})  
for the scalar form factor can be implemented easily. For the vector form factor it is useful to implement  also
the $P$-wave type  behaviour at the unitarity threshold $t_+$. As discussed in  \cite{BoCaLe}, 
this condition is equivalent to
\beq
\left[\frac{{\rm d} f_+(\tilde t(z, t_0))}{{\rm d} z}\right]_{z=1}=0, 
\eeq
 and amounts to the simple algebraic relation
\beq\label{eq:threshold}
M_{D^*} \Gamma_{D^*} \sum\limits_{j=0}^J b_{+,j} +2 (M_{D^*}^2-t_+) \sum\limits_{j=0}^J j b_{+,j}=0,
\eeq
 which must be fulfilled by the parameters $b_{+,j}$, simultaneously with the condition (\ref{eq:unitbn}).  These constraints are useful in the fits of the data, reducing the number of independent  parameters.

\bfig[ht]	
	\begin{center}\vspace{0.cm}
	 \includegraphics[angle = 0, clip = true, width = 2.3 in]
{Dpi.eps}	
	\end{center}\vspace{-0.2cm}
	\caption{Fit of a sample of data on the vector form factor $f_+(t)$ in the semileptonic region with the parametrization (\ref{eq:zparam+}). }
	\label{fig:fit}
\efig	

In order to illustrate the properties of the new parametrizations,  we generated from the CLEO data \cite{Ge:2008yi} a sample of values and errors for the vector form factor $f_+(t)$. We obtained the values  from Table XIII of  \cite{Ge:2008yi}, using for convenience $|V_{cd}|=0.236$ as in \cite{Ge:2008yi}. Assuming that the quoted values  correspond to the center of the bins defined in Table IX of \cite{Ge:2008yi}, and  adding $f_+(0)$ from LCSR as quoted in sect. \ref{sec:input}, we obtained the 10 data points given in Table \ref{table:coeff3}.  We then made a fit of these data using the expressions (\ref{eq:zparam+}), with the threshold condition (\ref{eq:threshold}), using both $t_0=0$ and $t_0=t_{opt}$.  The results of the two choices are actually very similar:  with 2 independent parameters $b_{+,0}$ and  $b_{+,1}$, the third one  $b_{+,2}$ being determined from eq.(\ref{eq:threshold}), we obtained for $t_0=0$ a minimal  $\chi^2=9.71 $ with the optimal parameters:
\beq\label{eq:bjt0}
  b_{+,0}= 0.665, \quad   b_{+,1}=0.725, \quad  b_{+,2}=-1.066, 
\eeq
 while for  $t_0=t_{opt}$ from (\ref{eq:t0opt}) we obtained  $\chi^2=9.67$ and the optimal coefficients
\beq\label{eq:bjtopt}
     b_{+,0}= 0.757,  \quad    b_{+,1}=0.397, \quad  b_{+,2}=-0.852.
\eeq
  We have  checked that the coefficients (\ref{eq:bjt0}) and (\ref{eq:bjtopt}) satisfy the 
constraint (\ref{eq:unitbn}) with $B_{jk}^{(+)}$ from (\ref{eq:Bij+}) and (\ref{eq:Bijtopt+}), 
respectively. The above parametrizations lead actually to almost identical form factors in the 
semileptonic region, as shown in Fig. \ref{fig:fit}. However, the choice  $t_0=t_{opt}$ allows a 
more accurate determination of the systematic error: assuming, as in \cite{BoCaLe}, that a reasonable 
definition of the truncation error is given by $|b_{+,3}|_{max} |\tilde z(t,t_0)|^3$, where   
$|b_{+,3}|_{max}$ is the maximum value of the next coefficient allowed by the condition 
(\ref{eq:unitbn}) at the optimal point, we find for $t_0=0$, when  $|b_{+,3}|_{max}=5.26$ 
and $\tilde z(t_-, 0)=0.35$, that this error reaches a value of about 7\%, while for   
$t_0=t_{opt}$, when $|b_{+,3}|_{max}=4.28$ and $\tilde z(t_-, t_{opt})=0.17$,  the truncation error can be reduced to less than  1\% in the whole semileptonic region. 

The nontrivial role played by the condition (\ref{eq:unitbn}) is seen if one makes a fit with one more 
independent parameter in (\ref{eq:zparam+}). Then, for  $t_0=t_{opt}$, the best fit without constraints
gives $\chi^2=6.86$. However, the unitarity condition (\ref{eq:unitbn})  is violated by the optimal parameters, 
and should be imposed explicitly for a suitable fit. 

We finally remark that although the parametrization (\ref{eq:zparam+}) for  $t_0=t_{opt}$ is defined such as to optimize the description of the semileptonic region, it has a reasonable behaviour also outside this region and even  on the unitarity cut. In particular,  the polynomial in the numerator  of (\ref{eq:zparam+}) becomes complex for $t>t_+$, when $|z|=1$, but one can check that the phase of the product in (\ref{eq:zparam+})  increases smoothly above the unitarity threshold and the resonance shape is not distorted, especially when the condition (\ref{eq:threshold}) is imposed. 

 The above exercise demonstrates the usefulness of the new parametrization (\ref{eq:zparam+}) for data analysis.
Of course, in a complete study the new expressions should be used directly in the fit of the partial branching fractions over bins in $t=q^2$,  given in \cite{Ge:2008yi}. The values provided by lattice calculations in the physical region \cite{arXiv:0903.1664,Bailey:2010vz,DiVita:2011py} can be also included in a combined fit, allowing a simultaneous determination of the vector form factor and of $|V_{cd}|$, as was done for the similar $B\pi$ case in \cite{BoCaLe}.

\section{Conclusions\label{sec:conc}}
 
In this work, we have explored the implications of analyticity and unitarity for the $D\pi$ form factors.
The work was motivated by the recent experimental measurements of the semileptonic $D\to\pi l\nu$ decay  
\cite{Ge:2008yi,Besson:2009uv} of interest for the extraction of the element $|V_{cd}|$ of the CKM matrix. 
We have applied the formalism of unitarity bounds, which use as input the dispersion relations satisfied by 
the moments of suitable heavy-light correlators at $q^2=0$. The available calculations of the moments 
in perturbative QCD up to $O(\alpha_s^2)$ terms \cite{Chetyrkin:2001je} allow more precise predictions 
in this framework. We have derived a family of constraints to be 
satisfied by the shape parameters of both the scalar and vector form factors at $t=0$ and explored the 
region on the real axis and in the
 complex plane where the form factors cannot have zeros.  The theoretical
input at the Callan-Treiman point and the phase information for the scalar form factor play an 
important role in constraining the corresponding shape parameters.  The exclusion regions for the zeros
that we have isolated basically cover a significant part of the entire low energy region. 

We have also proposed an improved parametrization of the $D\pi$ form factors in the semileptonic region, 
by properly implementing the singularities related to the lowest charmed resonances. 
The new parametrizations, given  in eqns. (\ref{eq:zparam+}) and  (\ref{eq:zparam0}) for the vector and 
scalar case, respectively, are based on truncated expansions in powers of the conformal mapping 
$z=\tilde z(t,t_0)$ defined in (\ref{eq:conformal}), with coefficients satisfying the quadratic condition 
(\ref{eq:unitbn}). Additional low-energy constraints or the threshold behaviour for the vector 
form factor can be easily implemented, as shown above. Using a sample of values produced from the CLEO experimental data  \cite{Ge:2008yi}, we demonstrated the usefulness of the new parametrization   for the description of 
the semileptonic data and a reasonable estimate of the systematic error. A more complex analysis, allowing an  accurate simultaneous  prediction of the form factor and of $|V_{cd}|$, implies  combined  fits  of the partial branching fractions over bins and of the LCSR and lattice predictions. This analysis will be reported in a future work.

\vskip0.5cm
{\small {\bf Acknowledgements:} 
BA thanks the Department of Science and Technology, Government of India, and the Homi Bhabha Fellowships 
Council for support. IC acknowledges support from  CNCS in the Program Idei, Contract No. 464/2009 and
from ANCS, Contract PN 09370102.}

\end{document}